\documentclass[10pt,letterpaper]{article}
\usepackage{opex3}
\usepackage{epstopdf}
\usepackage{cite}

\begin{document}

\title{Interference theory of metamaterial perfect absorbers}

\author{Hou-Tong Chen}

\address{Center for Integrated Nanotechnologies, Los Alamos National Laboratory, Los Alamos, New Mexico 87545, USA}

\email{chenht@lanl.gov} 



\begin{abstract}
The impedance matching in metamaterial perfect absorbers has been believed to involve and rely on magnetic resonant response, with a direct evidence from the anti-parallel directions of surface currents in the metal structures. Here we present a different theoretical interpretation based on interferences, which shows that the two layers of metal structure in metamaterial absorbers are linked only by multiple reflections with negligible near-field interactions or magnetic resonances. This is further supported by the out-of-phase surface currents derived at the interfaces of resonator array and ground plane through multiple reflections and superpositions. The theory developed here explains all features observed in narrowband metamaterial absorbers and therefore provides a profound understanding of the underlying physics. 
\end{abstract}

\ocis{(160.3918) Metamaterials, (230.5750) Resonators, (310.1620) Interference coatings, (310.6805) Theory and design.} 


\section{Introduction}

The demonstration of metamaterial perfect absorbers~\cite{Alu2005ICEAA,Landy2008PRL} represents one of the most important applications employing the astonishing properties found in metamaterials~\cite{Smith2000PRL,Chen2011LPR,Cai2010}. The original idea~\cite{Landy2008PRL} is that, in metamaterials with simultaneous electrical and magnetic resonances, both of the effective permittivity $\epsilon(\omega)$ and permeability $\mu(\omega)$ are highly dispersive and can be tailored independently. At certain frequencies the effective impedance, which is defined as $Z(\omega) = \sqrt{\mu(\omega)/\epsilon(\omega)}$, matches to the free space impedance $Z_0$, and therefore the reflection is minimized. If at the same time the metamaterial is also of high loss which causes very low transmission, then near-unity absorption can occur within an ultra thin layer of material. Under such considerations, in a typical metamaterial absorber, the magnetic resonance from the bi-layered metal structure~\cite{Zhang2005,Shalaev2005} is essential, and each layer of metal structure also provides the electrical response. As a direct evidence of the magnetic resonance in metamaterial absorbers, the surface currents excited in the two metal layers are found to be anti-parallel~\cite{Tao2008OE}. 

While all of the above arguments seem to be plausible and accepted by a majority of researchers~\cite{Avitzour2009PRB,Liu2010NL,Hao2011PRB}, fundamental questions remain. In fact, such a metamaterial absorber can be equivalent to a single layer of atoms (or molecules), and it is strongly inhomogeneous in the wave propagating direction~\cite{Zhou2011}. So it can hardly be considered as an effective bulk medium where the constitutive parameters $\epsilon(\omega)$ and $\mu(\omega)$ apply. Additionally, Fabry-P\'erot resonance~\cite{Zhou2011,Chen2010PRL, Sun2011OE}, conventional transmission line model~\cite{Wen2009OE}, and cavity resonance~\cite{Shchegolkov2010PRB} have been proposed to explain metamaterial absorbers or antireflection coatings, which also raise questions regarding the existence or involvement of the magnetic resonance in metamaterial absorbers. 

In this paper, we start with the multiple reflections interference model~\cite{Chen2010PRL}, and show in a typical metamaterial absorber there is negligible near-field interaction or magnetic response between the neighboring metal structures. The two layers of metal structures can be decoupled and the only link is multiple reflections between them. Through combination of numerical simulations and analytical calculations, we further derive the surface currents and reveal that the anti-parallel directions are a  result of interference and superposition, rather than excited by the magnetic component of the incident electromagnetic fields.

\begin{figure}[b!]
\centerline{\includegraphics[width=3in]{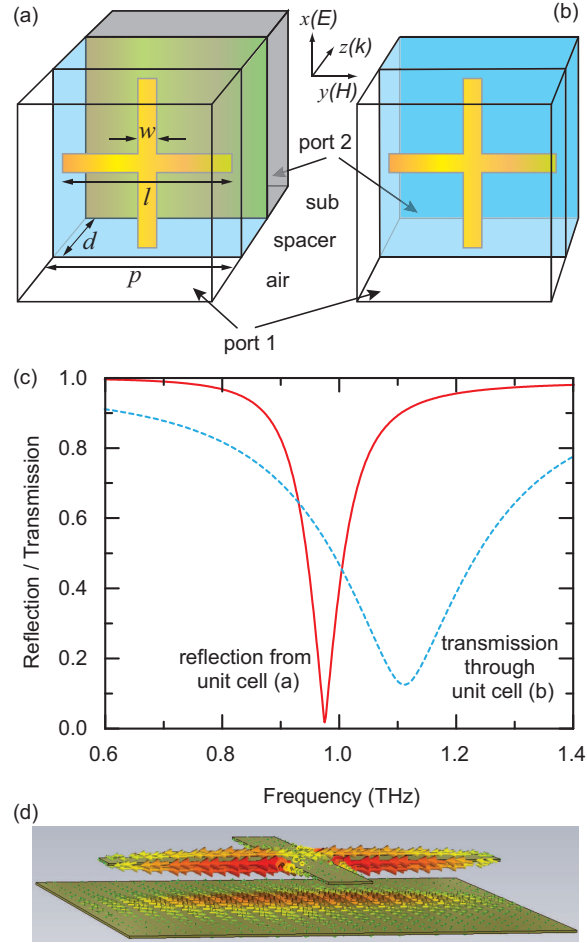}} \caption{(a) Schematic structure of the metamaterial absorber unit cell, which repeats in $x$ and $y$ directions forming a square array with periodicity of $p$. (b) Unit cell used to obtain the reflection and transmission coefficients at air-spacer interface with the cross-resonator array. (c) The solid red curve is the reflection $|S_{11}|$ from the metamaterial absorber shown in unit cell (a), and the dashed blue curve is the resonant transmission $|S_{21}|$ through the cross-resonator array shown in unit cell (b). (d) Surface currents of the metamaterial absorber at the peak absorption frequency. The current at the cross flows to left direction, while it flows to right direction at the ground plane. } \label{Fig1}
\end{figure}

\section{Models in numerical simulations}
\subsection{Coupled system}

Without losing generality, we use a simple and typical metamaterial absorber structure operating at terahertz frequencies as an example. It consists of a cross-resonator array and a ground plane separated by $d = 10~\mu$m thick polyimide dielectric spacer~\cite{Tao2008PRB,Diem2009PRB,Liu2010PRL}. The unit cell is shown in Fig.~\ref{Fig1}(a), where the periodicity is $p = 100$~$\mu$m, the length and width of cross wires are $l = 90~\mu$m and  $w = 10~\mu$m, respectively, and the thickness of metal is $t = 0.2~\mu$m. Due to the presence of the ground plane, the substrate provides the mechanical support only and the choice of material does not affect the performance of the metamaterial absorber. We carry out numerical simulations using CST Microwave Studio 2009~\cite{CST2009}. Perfect electric conductor (PEC) is used to simulate the metal, while the ohmic loss is effectively put in the polyimide spacer with dielectric constant $\epsilon_{\rm spacer} = 3.1\times(1 + 0.07 i)$. Periodic boundary conditions are applied to the side walls parallel to $z$ direction, and the ports are at the front and back surfaces of the unit cell. The the simulated electric field reflection $|S_{11}|$ is shown by the solid red curve in Fig.~\ref{Fig1}(c), which reveals a nearly zero reflection at a frequency slightly below 1~THz. This also suggests that the incident light is almost completely absorbed by the metamaterial in this narrow frequency band, according to  $A(\omega) = 1 - R(\omega) - T(\omega)$, where $R(\omega) = |S_{11}|^2$ is the reflectance, and $T(\omega) = |S_{21}|^2 = 0$ is the zero transmittance due the presence of the ground plane. The surface currents in the cross-resonator array and the ground plane exhibit anti-parallel directions, as shown in Fig.~\ref{Fig1}(d).

\subsection{Decoupled system}

The above model treats the metamaterial absorber as a {\it coupled} system, i.e., the possible near-field interactions and magnetic resonance have been taken into account between the cross-resonator array and the ground plane. However, we may also consider the presence of the metal cross-resonator array resulting in an impedance-tuned air-spacer interface~\cite{Holloway2009} with dramatically modified complex reflection and transmission coefficients shown in Fig.~\ref{Fig2}. The ground plane, on the other hand, functions as a perfect reflector with reflection coefficient $-1$. In the interference model we {\it decouple} the metamaterial absorber into two tuned interfaces, respectively with cross-resonator array and ground plane located at the two sides of the spacer. They are only linked by multiple reflections, as shown in the inset to Fig.~\ref{Fig2}(a), while any near-field interaction or magnetic resonance has been neglected. In this model, we need to know the reflection/transmission coefficients at the air-spacer interface with cross-resonator array, which are simulated using the unit cell shown in Fig.~\ref{Fig1}(b), where the ground plane and the substrate have been removed from the unit cell shown in Fig.~\ref{Fig1}(a). The cross-resonator array reveals a dipole resonance as indicated by the deep transmission dip at about 1.1~THz, shown in the dashed blue curve in Fig.~\ref{Fig1}(c). Comparing the two curves, the feature of particular interest and worthy of mentioning is that near-zero reflection (or unity absorption) of the metamaterial absorber occurs at a frequency different from and lower than the resonance frequency of the cross-resonator array. 

\section{Interference theory}

\begin{figure}[h!]
\centerline{\includegraphics[width=3in]{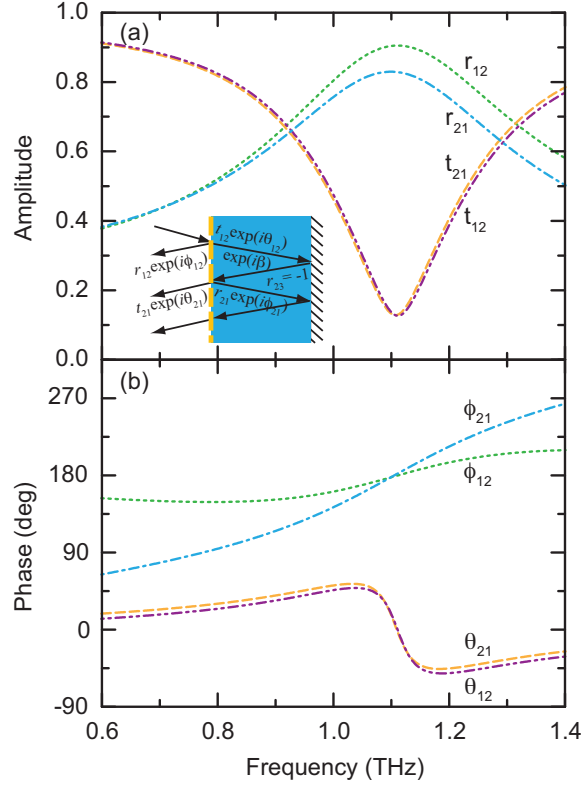}} \caption{(a) Amplitude and (b) phase of the reflection and transmission coefficients at the air-spacer interface with cross-resonator array, obtained by numerical simulations using the unit cell shown in Fig.~\ref{Fig1}(b). Inset: Multiple reflections and interference model of the metamaterial absorber, where the cross-resonator array is indicated by the dashed line at the air-spacer interface. } \label{Fig2}
\end{figure}

As shown in the inset to Fig.~\ref{Fig2}(a), at the air-spacer interface with cross-resonator array, the incident light is partially reflected back to air with a reflection coefficient $\tilde r_{12} = r_{12} e^{i \phi_{12}}$ and partially transmitted into the spacer with a transmission coefficient $\tilde t_{12} = t_{12} e^{i \theta_{12}}$. The latter continues to propagate until it reaches the ground plane, with a complex propagation phase $\tilde\beta = \beta_{\rm r} + i \beta_{\rm i} =\sqrt{\tilde\epsilon_{\rm spacer}} k_0 d$, where $k_0$ is the free space wavenumber, $\beta_r$ is the propagation phase, and $\beta_i$ represents the absorption in the spacer. After the reflection at the ground plane and addition of another propagation phase $\tilde \beta$, partial reflection and transmission occur again at the air-spacer interface with cross-resonator, with coefficients of $\tilde r_{21} = r_{21} e^{i \phi_{21}}$ and $\tilde t_{21} = t_{21} e^{i \theta_{21}}$, respectively. Similar to the light propagation in a stratified media~\cite{Born1980}, the overall reflection is then the superposition of the multiple reflections: 
\begin{equation}
\tilde{r} = \tilde r_{12} - \frac{\tilde t_{12} \tilde t_{21} e^{i 2 \tilde\beta}}{1 + \tilde r_{21} e^{i 2\tilde\beta}},
\label{Reflection}
\end{equation}
where the first term is the reflection directly from the cross-resonator array, and the second term, including the ``$-$'' sign, is the reflection resulting from superposition of the multiple reflections between the cross-resonator array and ground plane. The absorptance is then retrieved through $A(\omega) = 1 - |\tilde r(\omega)|^2$ since the transmission is zero. 
\begin{figure}[h!]
\centerline{\includegraphics[width=3in]{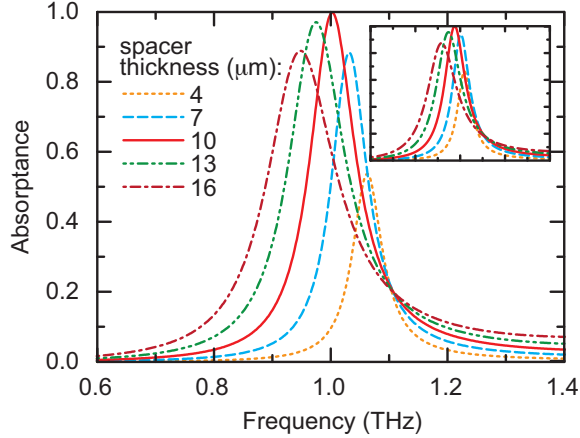}} \caption{Calculated absorptance in the decoupled metamaterial absorber using the interference model for various spacer thicknesses. Insets: The simulated absorptance when treating the whole metamaterial absorber as a coupled system.} \label{Fig3}
\end{figure}

\subsection{Spacer thickness dependent absorption}
Using the reflection and transmission coefficients shown in Fig.~\ref{Fig2}, the absorptance is calculated in the decoupled metamaterial absorber system and the results are shown in Fig.~\ref{Fig3}, where we have purposely varying the spacer thickness $d$ from 4~$\mu$m to 16~$\mu$m. The spacer thickness dependent absorptance in the coupled metamaterial absorber is also simulated directly using the unit cell shown in Fig.~\ref{Fig1}(a) and the results are shown in the inset to Fig.~\ref{Fig3}, which is in excellent agreement with the results calculated in the decoupled metamaterial absorber using the interference model. With all of these spacer thicknesses, we observe an absorption peak, which increases in the beginning and its frequency red-shifts as the spacer thickness increases. It reaches near-unity absorption at 1~THz when the spacer thickness is about 10~$\mu$m, and then the peak value decreases and its frequency continuously red-shifts when the spacer thickness further increases. Further numerical simulations and calculations reveal that, with any chosen spacer dielectric material, there is always an optimized spacer thickness where the reflection can reach zero, i.e. unity-absorption. 

The excellent agreement shown in Fig.~\ref{Fig3} validates the interference model we have presented above. The physical explanation of a metamaterial absorber is then as follows. The multiple reflections in the metamaterial absorber, i.e. the second term in Eq.~(\ref{Reflection}), constructively interfere as evidenced by the fact that near 1~THz the phase change of a round trip is $2 \beta_{\rm r} + \phi_{21} +180^\circ \approx 360^\circ$. The superposition of the multiple reflections then destructively interferes with the direct reflection from the air-spacer interface with cross-resonators, i.e. the first term in Eq.~(\ref{Reflection}). With the optimized spacer thickness, these two terms cancel each other out resulting in zero reflection; while with other spacer thicknesses the amplitude and phase do not match and these two terms only partially cancel each other out, resulting in a reduced absorption peak and corresponding frequency shift. Obviously, this explanation does not involve any near-field interaction or magnetic resonance between the two metal layers in the metamaterial absorber.

\subsection{Anti-parallel surface currents}

\begin{figure}[b!]
\centerline{\includegraphics[width=3in]{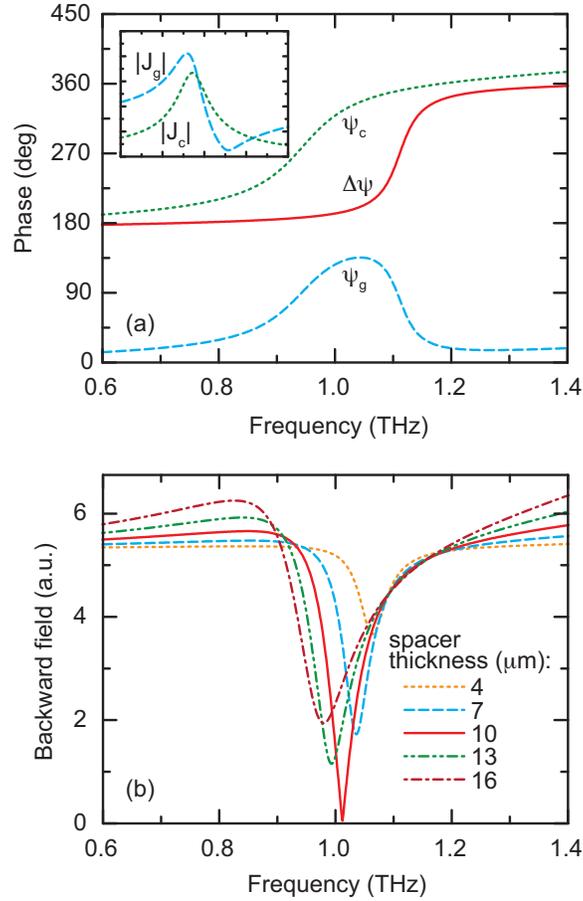}} \caption{(a) Phase spectra of the excited surface currents at the air-spacer interface with cross-resonator array ($\psi_{\rm c}$) and at the ground plane ($\psi_{\rm g}$), as well as their difference ($\psi_{\rm c} - \psi_{\rm g}$), calculated through multiple reflections and superposition based on the interference model. The corresponding magnitude of the surface currents are shown in the inset. (b) Backward radiation field (reflection) calculated through ${\bf E}_{\rm r} \propto {\bf J}_{\rm c}^{\rm total} + {\bf J}_{\rm g}^{\rm total} e^{i \beta_0}$ for various spacer thicknesses.} \label{Fig4}
\end{figure} 

The remaining question is then why we observe the anti-parallel directions of the excited surface currents as shown in Fig.~\ref{Fig1}(d), which has been used as the evidence of a magnetic resonant response~\cite{Tao2008OE} similar to fish-net metamaterials~\cite{Zhang2005,Shalaev2005}. In order to elucidate this mystic observation, by considering the boundary conditions at the interfaces we derive the excited surface currents based on interference and superposition. When the incident light with electric field ${E}_{\rm c}^{\rm i} \hat{\textbf{x}}$ is first reflected and transmitted at the air-spacer interface with cross-resonator array, according to the boundary condition requirements~\cite{Jackson1998}, it excites a surface current $\textbf{J}_{\rm c} = [(1 -\tilde r_{12}) / Z_0 -\tilde t_{12}/Z_{\rm s})] {E}_{\rm c}^{\rm i} \hat{\textbf{x}}$, where $Z_0$ and $Z_{\rm s}$ are impedances of the free space and spacer, respectively. The transmitted light continues to propagate and reflected by the ground plane, and the latter excites a surface current at the ground plane $\textbf{J}_{\rm g} = 2 \tilde t_{12} e^{i\tilde\beta} E_{\rm c}^{\rm i} / Z_{\rm s} \hat{\textbf{x}}$. The reflection from the ground plane propagates back to the air-spacer interface with cross-resonator array, where partial reflection and transmission occur which excites a surface current $\textbf{J}'_{\rm c} =  [(1 -\tilde r_{21})/Z_{\rm s}  - \tilde t_{21}/Z_0] \tilde t_{12} e^{i 2 \tilde\beta} (-1) E_{\rm c}^{\rm i} \hat{\textbf{x}}$. This procedure continues and the overall surface currents at the cross-resonator array and ground plane are then superpositions of multiple excitations:
\begin{eqnarray}
{\bf J}_{\rm c}^{\rm total} & = & \left \{\left (\frac{1 - \tilde r_{12}}{Z_0} - \frac{\tilde t_{12}}{Z_{\rm s}} \right ) - \frac{\left (\frac{1 - \tilde r_{21}}{Z_{\rm s}} - \frac{\tilde t_{21}}{Z_0} \right ) \tilde t_{12} e^{i 2 \tilde \beta}}{1 + \tilde r_{21} e^{i 2 \tilde \beta}} \right \} E_{\rm c}^{\rm i} \hat{\textbf{x}} \label{eqJc},\\
{\bf J}_{\rm g}^{\rm total} & = & \frac{2 \tilde t_{12} e^{i \tilde \beta}}{1+\tilde r_{21} e^{i 2 \tilde \beta}} \frac{1}{Z_{\rm s}} E_{\rm c}^{\rm i} \hat{\textbf{x}}. \label{eqJg}
\end{eqnarray} 

With a spacer thickness of $d = 10~\mu$m, the phase and magnitude of surface currents in Eqs. (\ref{eqJc}) and (\ref{eqJg}) are shown in Fig.~\ref{Fig4}(a) and its inset, respectively, which reveal a phase difference of $\Delta \psi = \psi_{\rm c} - \psi_{\rm g} = 192^\circ$ and comparable surface current magnitude at 1~THz. That is, they are almost out-of-phase, directly resulted from the interference rather than the magnetic resonance. We also notice that, in the calculated results shown in Fig.~\ref{Fig4}(a), at frequencies below 1~THz the phase difference is $\sim$$180^\circ$, while at frequencies above the resonance of the cross-resonator array (about 1.1~THz) the phase difference is nearly $360^\circ$. This agrees well with the results in numerical simulations when treating the whole metamaterial absorber as a coupled system. Finally, we use the obtained surface currents to calculate the overall backward radiation field (i.e. reflection) by ${\bf E}_{\rm r} \propto {\bf J}_{\rm c}^{\rm total} +{\bf  J}_{\rm g}^{\rm total} e^{i \beta_0}$ for different spacer thicknesses. The results shown in Fig.~\ref{Fig4}(b) reveal a reflection dip, in which the spacer thickness dependence of the reflection minimum and its frequency is again consistent with the absorption results shown in Fig.~\ref{Fig2}.

\section{Conclusion}
We have shown that it is {\it not} required in a metamaterial absorber to have simultaneous electric and magnetic responses, which have been considered as the foundation of metamaterial absorbers. We explicitly demonstrate that the assumed magnetic resonance plays a negligible role in the impedance matching of metamaterial absorbers to free space. In contract, it is the destructive interference between the direct reflection and the following multiple reflections that effectively traps light in the metamaterial absorbers and eventually causes the high absorption. Based on such an interference model, we derive the surface currents through multiple excitations and superposition, which clearly show the anti-parallel directions of surface currents in the resonator array and ground plane. This profound understanding of the underlying physics will undoubtedly provide valuable guidance in the future developments of more advanced metamaterial absorber based devices for microwave and photonic applications.   

\section*{Acknowledgements}
The author acknowledges the fruitful discussions with J. Zhou, Y. Zeng, L. Huang, A. K. Azad, J. F. O'Hara, W. J. Padilla, R. D. Averitt, and A. J. Taylor. We acknowledge support in part from the Los Alamos National Laboratory LDRD Program. This work was performed, in part, at the Center for Integrated Nanotechnologies, a US Department of Energy, Office of Basic Energy Sciences Nanoscale Science Research Center operated jointly by Los Alamos and Sandia National Laboratories. Los Alamos National Laboratory, an affirmative action/equal opportunity employer, is operated by Los Alamos National Security, LLC, for the National Nuclear Security Administration of the US Department of Energy under contract DE-AC52-06NA25396.

\end{document}